\documentclass[fleqn,10pt]{wlscirep}
\usepackage[utf8]{inputenc}
\usepackage[T1]{fontenc}
\title{Revealing spatial spillover effect in high-tech industry agglomeration from a high-skilled labor flow network perspective}

\author[1]{Chen Wang}
\author[2,*]{Lu Wang}
\author[2]{Yanbo Xue}
\author[3,*]{Ruiqi Li}
\affil[1]{School of Economics and Management, University of Science and Technology Beijing, Beijing 100083, China}
\affil[2]{Career Science Lab, Boss Zhipin, Beijing 100028, China}
\affil[3]{UrbanNet Lab, College of Information Science and Technology, Beijing University of Chemical Technology, Beijing 100029, China}

\affil[*]{corresponding.wanglu02@kanzhun.com, lir@mail.buct.edu.cn}

\begin{abstract}
Understanding the high-tech industrial agglomeration from a spatial-spillover perspective is essential for cities to gain economic and technological competitive advantages. Along with rapid urbanization and the development of fast transportation networks, socioeconomic interactions between cities have been ever-increasing, traditional spatial metrics are not enough to describe actual inter-city connections. High-skilled labor flow between cities strongly influences the high-tech industrial agglomeration, yet receives less attention. By exploiting unique large-scale datasets and tools from complex network and data mining, we construct an inter-city high-skilled labor flow network, which was integrated into spatial econometric models. Our regression results indicate that spatial-spillover effects exist in the development of high-tech industries in the Yangtze River Delta Urban Agglomeration region. Moreover, the spatial-spillover effects are stronger among cities with a higher volume of high-skilled labor flows than among cities with just stronger geographic connections. Additionally, we investigate the channels for the spillover effects and discover that inadequate local government expenses on science and technology likely hamper the high-tech industrial agglomeration, so does the inadequate local educational provision. The increasing foreign direct investments in one city likely encourages the high-tech industrial agglomeration in other cities because of the policy inertia toward traditional industries.
\end{abstract}
\begin{document}

\flushbottom
\maketitle

\section{Introduction}

Along with the rapid urbanization, cities have become the engines for economic and technological developments \cite{ref1}. Cities today are cooperating and competing, while the development of high-tech industries has become core strategies to help a city gain economic and technological competitive advantages \cite{ref2}, as the knowledge-intensive high-tech industry has been remaining at the forefront of manufacturing industries in technological innovation \cite{ref3}. In particular, compared with traditional industries, high-tech industries are more likely to agglomerate because their development is less constrained by natural geographical conditions \cite{ref4}.

A significant amount of research effort has been made to explain the agglomeration of high-tech industries from a spatial spillover perspective. A spatial spillover effect is defined as the fact that the degree of industrial agglomeration in a region depends on some influencing factors from another \cite{ref5}. For example, local governments tend to create policy environments that can promote the development of high-tech industries in their jurisdictions
\cite{ref6,ref7}. This will in turn cause other regions, especially the neighboring ones, to put forth similar policies, which will consequently affect the degree of industrial agglomeration in the imitated region \cite{ref8}. It is worth noting that the extent to which spatial spillovers can be described in other new forms expect for geographic distance is not well established in both theoretical and empirical research considering most studies have long been using geographic weight matrices, {\em e.g.}, Euclidean distance, length of adjacent boundary, to measure connections between adjacent regions.

However, the essence of ``spatial connections" keeps being updated. When the transportation network is not that fast and convenient, the geographic proximity plays an important and dominant role; even some deadly epidemic can only spread as a diffusion process with a speed of 300-600km/year \cite{ref7,ref9}. Yet in modern times, fast transportation networks (including highways, airlines, rapid trains) have twisted the Euclidean space into a higher dimensional space, where proximity cannot be fully explained by traditional Euclidean distance. For example, Beijing and Shanghai are far from each other measured by geometric distance, yet with fast transportation and high volume of passenger flow, Beijing is much closer to Shanghai than to a small mountain village even adjacent to Beijing. In this sense, a weighted topological space defined by labor flow is more suitable to describe the real connection between cities when studying industry agglomeration. More than this, the frequent labor flow caused by transportation advancements has established close links between non-adjacent regions. The decisive role of labor flows in clustering formation attributes to labor market pooling and knowledge spillover \cite{ref10}. As empirically tested, high-skilled individuals are more internally mobile \cite{ref11} and tend to move to places where more firms and similar jobs are provided, forming a skill- and knowledge-concentrated local community \cite{ref12}. Besides, the complementarities of proximate skills will cross different but related industries over the long run \cite{ref13}, allowing to optimize the structure and quality of the entire industrial workforce through technology spillovers \cite{ref14}.

Therefore, it is essential to break through the limitation of geographical proximity and consider the spatial spillover effect caused by labor flows on the agglomeration of high-tech industries. Such flow connection in the form of high-skilled migrants is in line with the trend of social development, which is helpful for designing policies on developing the high-tech industries. However, a challenge is how to construct an alternative socioeconomic matrix, along with the geographical measurements to comprehensively describe spatial effects in a weighted topological flow space, which has not been discussed in the current literature on the influencing factors of the agglomeration of high-tech industries \cite{ref15}. This paper aims to answer the following three key questions:

(1) How to reveal the flow connection caused by high-skilled labor migration between cities?

(2) What kind of spatial spillover effect does the above flow connection have on the agglomeration of regional high-tech industries?

(3) How is the spatial spillover effect caused by the above flow connection different from traditional geographic connection?

\begin{figure}
  \centering
  \includegraphics[width=0.5\linewidth]{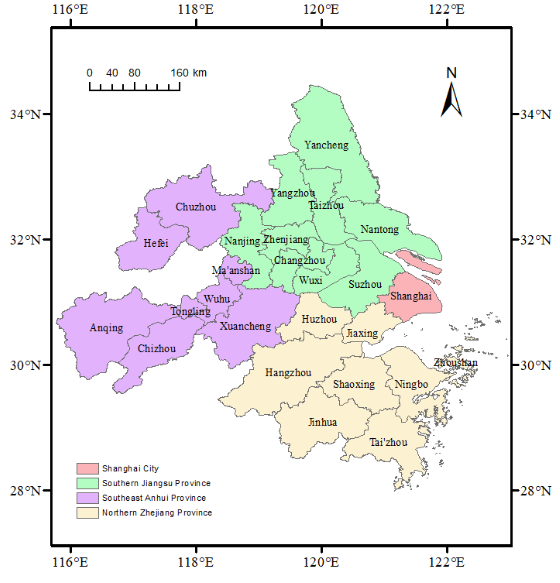}
  \caption{The 26 cities of the Yangtze River Delta Urban Agglomeration (YRDUA) region.} 
  \label{fig1}
\end{figure}

\begin{figure}
  \centering
  \includegraphics[width=\linewidth]{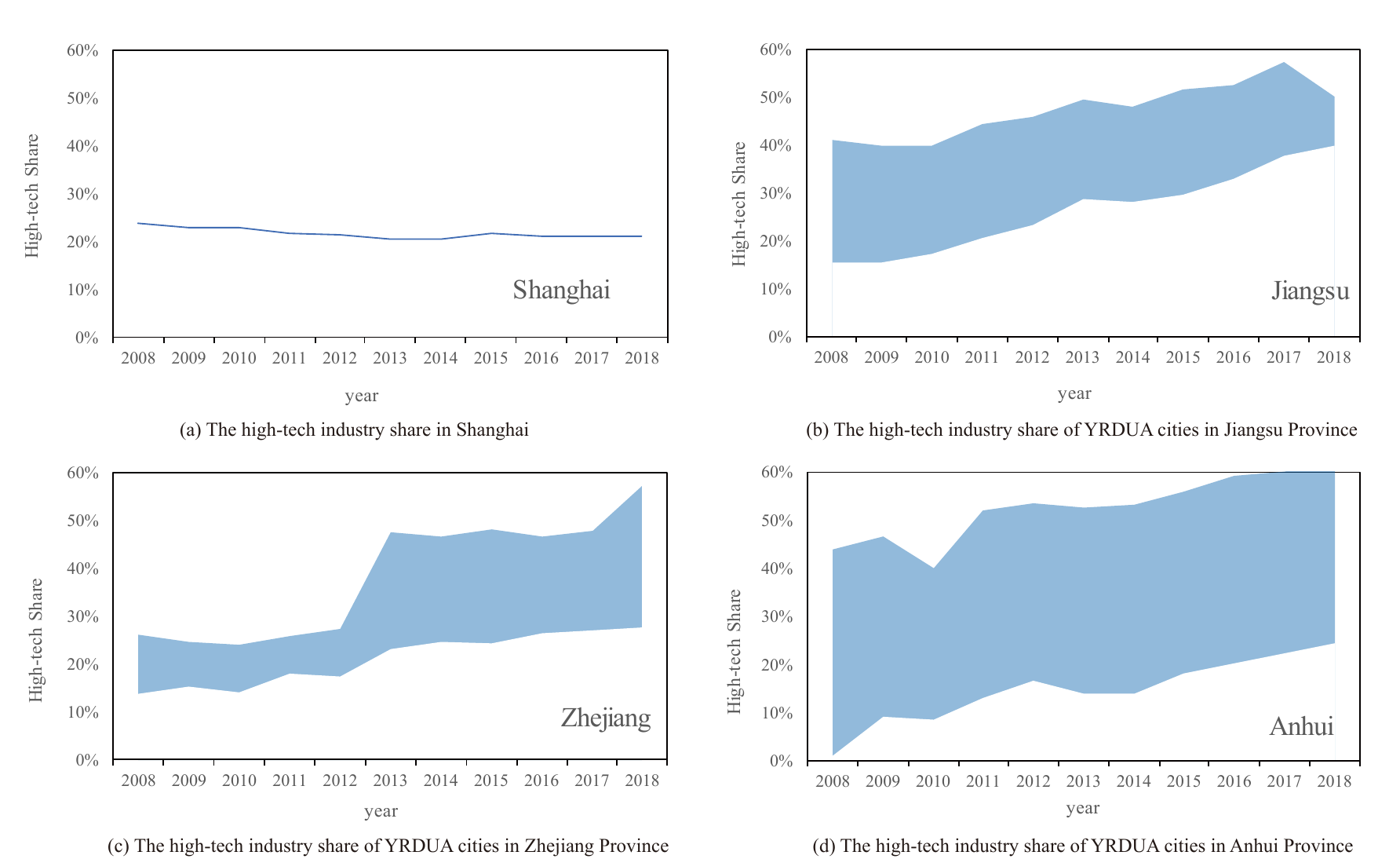}
  \caption{Min-max share of high-tech manufacturing industries of cities in YRDUA from 2008 to 2018. 
}
  \label{fig2}
\end{figure}

To answer above questions, we take the Yangtze River Delta Urban Agglomeration (YRDUA) as an example. The YRDUA region includes 26 cities, with a land area of 211.7 million square kilometers, accounting for approximately 2.2\% of China's total area (Figure \ref{fig1}). As the core base of high-tech manufacturing industries in China, the YRDUA envisages a blueprint of advanced manufacturing industries and has engaged collaborative efforts in industrial upgrading through various internal supporting initiatives and extensive international cooperation \cite{ref16,ref17}. In 2018, the YRDUA had over 40 percent in the shares of advanced manufacturing industries, twice as that in 2008. Nevertheless, there is a distinct diverging trend across cities. Figure \ref{fig2} demonstrates the yearly progress in the development of high-tech industries measured by the share of high-tech manufacturing sectors relative to the above-scaled industries in terms of the gross output value, at both micro- and macro-regional levels in the last decade. Specifically, Shanghai, the center of the YRDUA, has a much slower progress in developing high-end manufacturing industries, whereas the role of high-tech industries in Anhui Province has been substantially promoted despite a distinct internal disparity within the YRDUA. Note that in YRDUA, Jiangsu is specific to its south area, Zhejiang is specific to the northern area, and Anhui denotes its southeast area. The data of the YRDUA is represented by the average of its 26 member cities in Figure \ref{fig2}.

Taking all above factors into consideration, this study validates the spatial spillover effect in a labor-flow space of macroeconomic factors that might influence the agglomeration of high-tech industries in the YRDUA using spatial econometric models. By exploiting a unique dataset and tools from big data mining \cite{ref18}, a new spatial weight matrix based on inter-city high-skilled labor flow is constructed and integrated into the spatial econometric model, which is characterized as a new measurement that facilitates regional connection.

The remainder of this paper is organized as follows. Section 2 reviews the related literature and summarizes the challenges of existing researches. Section 3 introduces three spatial econometric models and the data construction process. Section 4 discusses the regression results. Conclusions are provided in Section 5.

\section{Literature Review}

\subsection{Determinants of industry agglomeration}

Industry agglomeration refers to the concentration of economic activities from related firms in a geographical area \cite{ref19}. As an important topic in economic geography, the determinants of industry agglomeration have been discussed in many studies \cite{ref20}. Marshall's theory summarized that three factors, namely specialized skilled labor, the development of subsidiary trade and suppliers of intermediate inputs, and knowledge spillover, would drive industry agglomerations \cite{ref21}. Takeda {\em et al.} \cite{ref22} analyzed the relationship between geographical agglomeration and modularized industrial networks in Yamagata, Japan. They suggested that transportation infrastructure is the key to site location for firms due to the apparent fact that physical proximity can reduce transportation costs and provide easy access to other partners within their inter-firm network. Lu {\em et al.} \cite{ref23} investigated the determinants of China's manufacturing industries with a particular focus on local protectionism. The results indicate that local protectionism obstructs the process of geographic concentration of manufacturing industries. Song {\em et al.} \cite{ref24} examined the relationship between industry agglomeration and transportation accessibility within the Seoul Metropolitan area. They find that transport networks have exerted a positive impact on industry agglomeration, but the magnitude and significance of regression results varied across different industries. Akkemik {\em et al.} \cite{ref25} confirmed the explanatory power of new trade theory to industrial concentration. Lu {\em et al.}  \cite{ref26} found that fiscal support and higher human capital are the two main drivers of industry agglomeration. Imaizumi {\em et al.} \cite{ref27} investigated the persistent impact of the temporary shock of an earthquake on spatial industry agglomeration and found that the earthquake damaged old industrial clusters, which motivated non-damaged neighboring regions to attract both industries and talents to move.

\subsection{Impact of high-skilled labor migration on industrial agglomeration}

Since Lucas \cite{ref28} proposed that spillover effect can also exist in human capital, an increasing number of studies have generally acknowledged that urbanization promotes labor mobility and that large-scale population aggregation is a driver of knowledge exchange and knowledge spillover \cite{ref29}, and it also facilitates the accumulation, dissemination, and innovation of knowledge \cite{ref30}. Feldman {\em et al.} \cite{ref31} argued that the geographical aggregation of individuals can reduce the cost of research as well as the cost of turning technology into productivity. In an investigation of the internal migration between two cities, the core city had a larger industrial share than the peripheral city, and labor mobility became the direct cause of urban industrial agglomeration \cite{ref32}. Gagliardi \cite{ref33} argued that innovation relies on external sources such as skilled immigration inflows because skilled migration can not only bring novel knowledge, but also enable the creation of innovative environment which will later on foster further human capital accumulation. Fassio {\em et al.} \cite{ref34} demonstrated that highly-educated migrants have a positive effect on innovation, but the effect differs across industries.

With no doubts, the concentrated labor flow networks can generally increase the economic scale in the area \cite{ref35}, enhancing the agglomeration effect in the destination area \cite{ref36}. However, the effects will turn to be negative when the market crowding effect is greater than the aggregation of home-market effects and cost of living effects \cite{ref37}. Based on the core-periphery vertical link model, the researcher observed a triangular closed loop between labor flow, industrial transfer, and regional development gap \cite{ref38}.

\subsection{Paradigm shift of urban studies from geometric space to flow space}

The continuous evolution of the definition of ``city" has made relevant research gradually shift from geometric space to flow space. As Batty \cite{ref39} proposed in his book ``The New Science of Cities", cities should be defined as being a ``space of flows" rather than ``a place in space", and flows represents the strength of the connection between cities. For example, such a flow perspective is important on systematical traffic engineering, for which both the flow between locations and topology of transportation networks matter \cite{ref40}. Batty believes that to understand space, we must understand flow. To understand the urban issues from the perspective of flow seems have more significance in today's context that population mobility and information exchange are much more frequent. In particular, the diffusion of knowledge, epidemics, opinions and innovations are thought to have complex spatio-temporal patterns. In recent years, there was a research paradigm shift from geometrical space to flow space. The research published in Science by Brockmann and Helbing \cite{ref9} defined an effective distance based on human mobility flow network to successfully and accurately predict the arrival time of epidemic spread from the origin city. In comparison, the traditional geometric distance fails to make accurate prediction on the spreading process. Following this idea and in order to explore the complex epidemic spreading dynamics influenced by spatial structure and human dynamics, Li {\em et al.} \cite{ref10} integrated human mobility, human interaction intensity and demographic features to describe more realistic spatiotemporal patterns of epidemics, and discovered that as long as the distance is defined based on human mobility, the prediction results are quite comparable. Wesolowski {\em et al.} \cite{ref41} integrated more details of the epidemic (such as dengue) and predicted the geographic spreading patterns and first arrival timing of epidemics based on mobility data of travelers.

Also, there has been a few works focusing on the spatial spillover effects on industrial agglomeration in a flow space \cite{ref42,ref43}, as a single use of geographical connection is far from being enough to characterize the organic and emergent nature and dynamics of activities within regions \cite{ref11}. However, few studies have explored the determinants of industrial agglomeration from the perspective of flows. This is partially limited by the availability of empirical data on labor flows of specific occupations and locations \cite{ref19}. To address the aforementioned limitations, this paper describes regional connections in the form of industry-specific talent migration, allowing the dynamics to be better analyzed to help municipal governments formulate relevant policies for advanced manufacturing industries, and promote the regional integration and economic development.

\section{Method and Data}

\subsection{Basic model}

The local entropy index is often used to reflect the level of industry agglomeration because it eliminates regional size differences and can characterize the spatial distribution of geographical elements more realistically \cite{ref44,ref45}. The location entropy index is formed as follows:
\begin{eqnarray}\label{GrindEQ__1}
{Location\ Entropy}_{it}=\frac{e_{ih}/e_i}{\sum_i{e_{ih}}/\sum_i{e_i}},
\end{eqnarray}
where variable $e_{ih}$ denotes the gross output of high-tech industries in city \textit{i} in year \textit{t}. $e_i$ denotes the gross industrial output in city \textit{i} in year \textit{t}. The higher the location entropy index, the higher the agglomeration level.

The basic model is set as follows:
\begin{eqnarray} \label{GrindEQ__2}
{Location\ Entropy}_{it}&=&{\beta }_0+{\beta }_1{\mathrm{ln} {pcgdp}_{it}\ }+{\beta }_2{govexp}_{it}+{\beta }_3{fdi}_{it}+{\beta }_4{edu}_{it} +{\beta }_5t{radeopen}_{it}\nonumber\\ &+&{\beta }_6{\mathrm{ln} {finload}_{it}\ }+{\beta }_7{\mathrm{ln} {fixinvest}_{it}\ }+{\beta }_8{\mathrm{ln} {poweruse}_{it}\ }+u_i+{\varepsilon }_{it}. 
\end{eqnarray}

For each city\textit{ i,} \textit{Location Entropy} denotes high-tech industry agglomeration. \textit{pcgdp} denotes per-capita GDP; \textit{govexp} denotes the share of government expenditure on science and technology; \textit{fdi} is the ratio of inward foreign direct investments over GDP; \textit{edu} is the number of universities; \textit{tradeopen} denotes the openness degree measured by the proportion of the value of import and export trade in relation to GDP; \textit{finload} denotes total loads from non-back financial intermediaries at the end of the year; \textit{fixinvest} is the fixed asset investment; and \textit{poweruse} denotes the total electricity consumption. $u_i$ is the individual fixed effect, and $\varepsilon_{it}$ is the standard error term.

\subsection{Spatial econometric models}
Following the aforementioned discussion, the spatial effect of high-tech industry agglomeration and its influencing factors between cities likely exist. The spatial effect in spatial econometrics represents itself as the spatial autocorrelation of variables between different regions, which means that each variable in a region is related to that in its neighboring regions. In other words, these variables are spatially interdependent.

Considering the spatial interaction effects among the variables, three models can be used to explore the relationship between dependent and independent variables: the spatial lag model (SLM), spatial error model (SEM) and the spatial Durbin model (SDM; \cite{ref46}), each of which will be elaborated separately in the following:

In Eq. \eqref{GrindEQ__3_}, the SLM only contains the first type of spatial interaction effects,
\begin{eqnarray} \label{GrindEQ__3_}
{Location\ Entropy}_{it}=\rho W{(Location\ Entropy}_{it})+X_{it}\beta +u_i+{\varepsilon }_{it},
\end{eqnarray}
in which ${Location\ Entropy}_{it}$ denotes the dependent variable, $X_{it}\ $denotes the independent variables, and it is an $n\times k$ matrix. $\beta$ is a $k \times 1$ vector of the independent variable coefficient, $\rho$ represents the spatial regression correlation coefficient, and $W$ is the $n \times n$ spatial weights matrix. $W{(Location\ Entropy}_{it}\mathrm{)}$ is the spatial lag dependent variable, which is used to measure the spatial effects of the neighbor regions, and $\varepsilon_{it}$ is a random disturbance term. $n$ represents the number of observations, and $k$ represents the number of independent variables.

The SEM contains only the third type of interaction effects among the error terms $\varepsilon_{it}$ and is described as Eq. \eqref{GrindEQ__4_},
\begin{eqnarray}
{Location\ Entropy}_{it} &=&X_{it}\beta +u_i+{\varphi }_{it}, \nonumber\\
{\varphi }_{it}&=&\lambda W{\varphi }_{it}+{\varepsilon }_{it},\label{GrindEQ__4_}
\end{eqnarray}
where $\lambda$ represents the spatial regression error coefficient, and ${\varphi }_{it}$\textit{ }denotes the spatial autocorrelation error term.

The SDM includes both the first and second types of spatial interaction effects. Compared with SLM and SEM, this model is more comprehensive and flexible \cite{ref47}. In Eq. \eqref{GrindEQ__5_}, the meaning of $W{(Location\ Entropy}_{it})$ is the same as that in Eq. \eqref{GrindEQ__3_}. $\delta $ is the spatial autocorrelation coefficient of the independent variable, and when $\delta =0$, the SDM degenerates into SLM; when $\delta +\rho \beta =0$, the SDM degenerates into SEM.
\begin{eqnarray} \label{GrindEQ__5_}
{Location\ Entropy}_{it}=\rho W{(Location\ Entropy}_{it})+X_{it}\beta +WX_{it}\delta +{\varepsilon }_{it}.
\end{eqnarray}

As to how to choose the three spatial econometric models, LeSage \textit{et al.} \cite{ref46} asserted that ignoring the spatial interaction effects between the dependent variables and the independent variables produces more estimated bias than ignoring the spatial effects of error terms. Therefore, the SDM should be given priority \cite{ref47}. Next, it can be verified whether the SDM can be simplified into the SLM or SEM through the Wald test, likelihood ratio test or Lagrange multiplier test, and the three tests are approximately equivalent.

\subsection{Spatial weight matrix}

To analyze the spatial effects between variables, the spatial distance between the regions should be measured first. Denoting spatial distance between region \textit{i} and region \textit{j} as $W_{ij}$, the spatial weight matrix can be defined as follows:
\begin{eqnarray} \label{GrindEQ__6_}
W=\left( \begin{array}{ccc}
W_{11} & \dots  & W_{1n} \\
\dots  & \  & \dots  \\
W_{n1} & \dots  & W_{nn}\end{array}
\right).
\end{eqnarray}

In this paper, we set up three spatial weight matrices: weights based on adjacency principle, combined distance-boundary weights, and weights based on the migration of high-skilled talent. Based on these three matrices, three spatial models are proposed to explore the influencing factors of industrial agglomeration considering spatial spillover effects, the formation of each represents geographical proximity, overlapping boundaries, and high-skilled talent migration, respectively.

\subsubsection{Contiguity-based spatial weights}

This method constructs a space weight matrix based on whether two cities are adjacent. The set of boundary points of city \textit{i} is denoted by $bnd(i)$; then, the so-called queen contiguity weights are defined by Eq. \eqref{GrindEQ__7_}. However, the queen contiguity weight matrix allows for an extreme case, that is, the two cities share only a single boundary point. Hence, a stronger condition is required to more accurately measure the geographic connection between two cities.
\begin{eqnarray} \label{GrindEQ__7_}
w_{ij}=\left\{ \begin{array}{c}
1,\ bnd(i)\cap bnd(j)\neq \emptyset\\
0,\ bnd(i)\cap bnd(j)=\emptyset \end{array}
\right..
\end{eqnarray}

\subsubsection{Distance-boundary-based spatial weights}

In many situations, the spatial impact may exhibit aspects of not only boundary relations but also distance. Cliff \textit{et al.} \cite{ref48} proposed the space weight matrix by combining power distance and boundary shares, described in Eq. \eqref{GrindEQ__8_}. $l_{ij}$ denotes the length of the shared boundary between two cities. $d_{ij}$ is the centroid distances from each spatial unit \textit{i} to unit \textit{j} ($j\neq i\mathrm{)}$. 
\begin{equation} \label{GrindEQ__8_}
w_{ij}=\frac{l_{ij}d^{-1}_{ij}}{\sum_{k\neq i}{l_{ik}d^{-1}_{ik}}},\ \ \ i=1,2,\dots ,26.
\end{equation}

\subsubsection{Weights based on the high-skilled migration}

To analyze the spillover through population mobility between cities, a space weight matrix based on population mobility is constructed. Here, we use the inverse distance matrix in Eq. \eqref{GrindEQ__9_}. $s_{ij}$ is the number of the high-skilled talent migrating from city \textit{i} to city \textit{j}. Consistent with the literature, the matrix is normalized according to row standardization to interpret the spatial spillover effects as an average of all neighboring cities. Data on population mobility between cities are explained in detail in Section 3.5.
\begin{equation} \label{GrindEQ__9_}
w_{ij}=\frac{s_{ij}}{\sum_{k\neq i}{s_{ik}}},\ \ \ i=1,2,\dots ,26.
\end{equation}

\subsection{Global spatial autocorrelation}

Spatial autocorrelation can be understood as follows: regions with similar locations have similar variable values. To test the spatial dependence of high-tech industry agglomeration, we use the global Moran's \textit{I} index to perform the calculation. The Moran's \textit{I} index is formulated as Eq. \eqref{GrindEQ__10_}:
\begin{eqnarray} \label{GrindEQ__10_}
\textrm{Moran's}\ I=\frac{n\sum^n_{i=1}{\sum^n_{j=1}{w_{ij}\left(x_i-\overline{x}\right)\left(x_j-\overline{x}\right)}}}{S_0\sum^n_{i=1}{{\left(x_i-\ \overline{x}\right)}^2}},
\end{eqnarray}
with $S_0=\sum^n_{i=1}{\sum^n_{j=1}{w_{i,j}}}$. In Eq. \eqref{GrindEQ__10_}, $x_i$ and $x_j$ represent the location entropy of city $i$ and city $j$, respectively, and $w_{ij}$ denotes the spatial weight matrix. The value range of Moran's $I$ index is $[-1, 1]$. If a city with high value is adjacent to another city with high value, or a city with low value is adjacent with another city with low value, positive spatial autocorrelation is achieved, represented by a positive value of Moran's $I$ index; otherwise, if a city with a high-value variable is adjacent to a city with low value, negative spatial autocorrelation is achieved, represented by a negative Moran's $I$ index. A value closer to 1 or -1 indicates the significance of the spatial correlation or the spatial difference. Additionally, a value of 0 suggests no spatial correlation among variables, or the high-tech industry agglomeration presents a randomly spatial distribution.

\subsection{Data source}
To examine the macroeconomic determinants and their spatial spillovers of high-tech manufacturing industries, this study constructs a balanced panel, consisting of 26 cities in the YRDUA over the period from 2008 to 2018. Raw data regarding macroeconomic factors are gathered from the Science and Technology Department of Zhejiang\footnote{ http://www.zjkjt.gov.cn/html/node18/list3\_170402/170402\_1\_curM\_tt.html}, Jiangsu\footnote{ http://www.jssts.com/Category\_30/Index.aspx}, and Anhui Provinces\footnote{ http://www.ahkjt.gov.cn/content/channel/58774058ceab06625682f6b5/}, and the Shanghai Statistical Yearbook on Science and Technology (2009--2018). The data of 2018 are estimated by the annual growth rate recorded by the Provincial Statistical Yearbook and Reports on the Implementation of the 2018 Plan for National Economic and Social Development.

In this study, the high-skilled migrants are those who are employed workers or job seekers who expect to get a job in the high-tech manufacturing sectors including medicine, electronic chemicals, aircrafts and spacecrafts, computers and office equipment, medical equipment and measuring instruments, electronic equipment and communication equipment.

The data for building the weight matrix based on high-tech talents are retrieved from Boss Zhipin, one of the largest online recruiting platforms in China. By the end of 2018, it has recorded an accumulation of over 63 million online registers\footnote{Records are as of December, 2018.} identified as both employers and job seekers since its establishment in 2014. The wide adoption of service from Boss Zhipin makes it the best representative for analyzing the migration of high-skilled workers.

A job seeker's profile includes names and locations of his/her previous and current working places. The actual movement of job seekers occurred when the location of their working city changes, which is the proxy for change of residency. For the most part, the data of a company's location can be directly accessed from the database. Only those who have filled information or moved to one of the sampled cities within the YRDUA in a specific year are included in our datasets. The cases of intra-organizational mobility across different cities are also included. Registered samples with no records of moving to another city, or jumping to a new organization in the same city in a specific year, are all excluded from our datasets. However, on rare occasions, new information of working locations is missing, thus, we took a job seeker's expected destination as a replacement. This only accounts for less than 1 percent of the samples.

To ensure the flow data comparable with other variables, the monthly migratory flows are aggregated annually. In this way, we have gathered a unique large dataset including micro-level information of individuals distinguished by occupations, which allows us to conduct a nuanced analysis of inter-regional behaviors. More details of data processing methods can be found in \cite{ref19}.

\section{Results and Discussions}

\subsection{ Flow analysis}

\begin{figure}[htbp]
  \centering
  \includegraphics[width=0.5\linewidth]{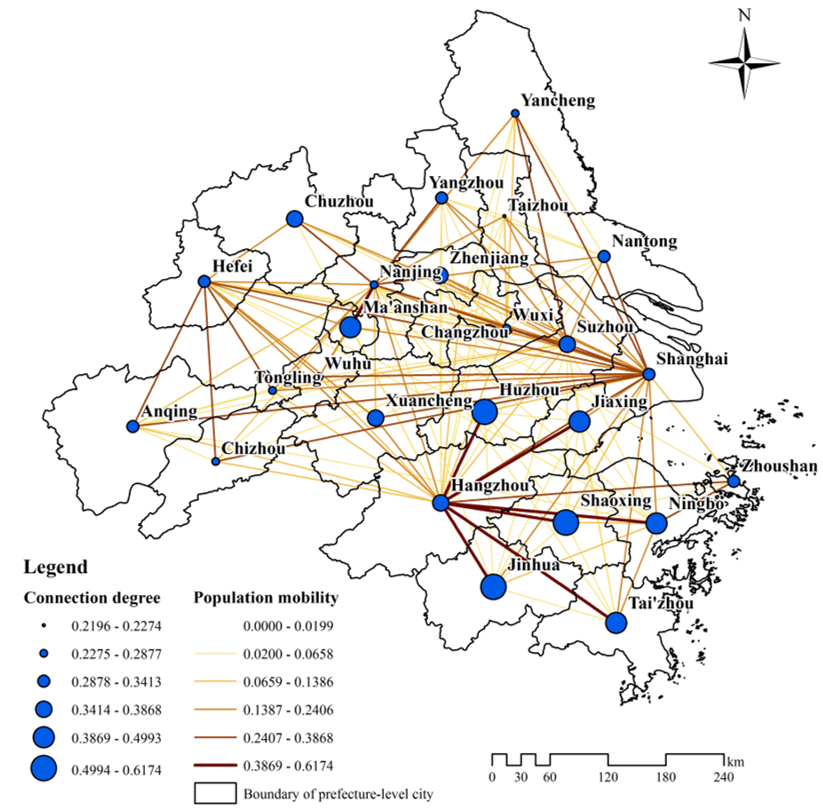}
  \caption{Migration flow between cities in YRDUA region (Data source: Boss Zhipin, 2015-2018).}
  \label{fig3}
\end{figure}

To visualize the talent flow, Figure \ref{fig3} plots the intensity of high-skilled talent migration between cities in the YRDUA region. As shown in Figure \ref{fig3}, Shanghai, Hangzhou, Nanjing, Suzhou and Hefei are regions where migrants flows are relatively concentrated. Among all, Shanghai, as the top central node of migration, has the largest intensity than other cities. An obvious reason is that its developed economy and abundant innovation resources are more appealing to people seeking personal development. Recent years have witnessed the rising of another central node in the YRDUA region, namely Hangzhou, reflected by Figure \ref{fig3}. It is worth to note that Shanghai and Hangzhou have different migration-receiving patterns, in that Shanghai tends to indiscriminately attract talents from the entire YRDUA region, but high-skilled talents moving to Hangzhou are mostly from its neighboring cities, such as Shaoxing, Jinhua, Huzhou and Jiaxing. Within Jiangsu Province, both Nanjing and Suzhou act as the center nodes for high-skilled talents. Relatively speaking, for these two cities, the migration intensity of Nanjing is greater because it can attract talents not only from its parent province, but from spatially proximate cities in Anhui Province. Since Suzhou is close to Shanghai and has rapid economic growth in recent years, it plays a diversion role in attracting talents that spilled-over from Shanghai. Interestingly, although Nanjing and Suzhou are adjacent to each other, migration intensities between them are relatively high, which further consolidates our previous assumptions that non-spatial factors like information flow may play decisive roles in high-skilled talent migrations. Lastly, Hefei, as the provincial capital city of Anhui Province, becomes a rather weak central node, which is only attractive to talents within the same province.

\subsection{Spatial autocorrelation test}

\begin{figure}
  \centering
  \includegraphics[width=0.5\linewidth]{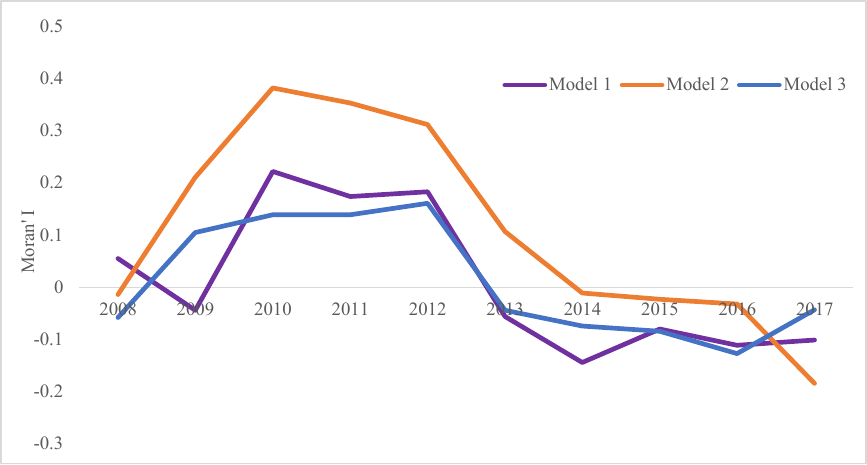}
  \caption{Moran's $I$ index values in three models. 
}
 \label{fig4}
\end{figure}

As discussed in Section 3.4, the extent of the spatial autocorrelation is tested by the global Moran's $I$ index. After calculating the location entropy value for each city, we obtain the Moran's $I$ index value for each year from 2008 to 2017 (see Figure \ref{fig4}). Note that Model 1-3 represent three spatial Durbin models based on three different types of spatial weight matrix, namely contiguity-based spatial weights, distance-boundary-based spatial weights, and weights based on the high-skilled migration, respectively. As we can see that the Moran's $I$ index values of the three spatial econometric models show trends from positive to negative, which indicates that the high-tech industry agglomeration in YRDUA exhibits positive spatial autocorrelation in earlier years and turns negative later. While the global Moran's $I$ index can reflect the average correlation level of the region as a whole, to individual cities, the positive spatial autocorrelation in some cities may be offset by the negative spatial autocorrelation in other cities. The relatively high Moran's $I$ index motivates us to further study the significance of all three different types of connections, as spatial autocorrelation cannot be interpreted as causality or wouldn't even guarantee a high significance.

\subsection{Spatial econometric regression results}

The regression results for three spatial econometric models are reported in Table \ref{tab1}. The global spatial coefficient $\rho$ is statistically significant at the 5\% level, indicating that high-tech industry agglomeration has a clear spatial dependence between neighboring cities. This result is similar to some of the views of the new economic geography, which states that there is a significant spatial spillover effect in the high-tech industry agglomeration within the YRDUA region \cite{ref18}.

Notably, the spatial spillover effects of the high-tech industry agglomeration do not exist in all adjacent cities because the global spatial coefficient $\rho$ is not significant in Model 1 where the spatial weights matrix is constructed simply by an adjacent principle. Only when the distance between the centers of adjacent cities is shorter and the boundary connectivity is stronger does the high-tech industry agglomeration reflect significant positive spatial dependence, that is, an increase in the high-tech industry agglomeration of neighboring cities promotes the high-tech industry agglomeration in this city.

A spatial weight matrix to reflect the flow connection caused by labor flow between cities is constructed in Model 3. Breaking though the scope of new economic geography, Model 3 considers whether the migration of high-skilled talent between two cities can lead to spatial spillover effects on high-tech industry agglomeration. The developed railway and highway transportation networks in the YRDUA region make the migration more convenient, which leads to the transfer of knowledge and technology. Therefore, the spatial spillover effects on high-tech industry agglomeration may exist not only between adjacent cities but also between cities with convenient talent flows. This spatial spillover effect is statistically proven in Model 3. The global spatial coefficient $\rho$ of Model 3 is statistically significant at the 1\% level, indicating that an increase in high-tech industry agglomeration in a city will be a driver of the agglomeration of high-tech industries in other cities which may not be adjacent but have frequent migration of high-skilled talents. By comparing the results of Model 3 with Model 2, it can be seen that spatial spillover effects between cities with flow connection are stronger and more significant than the ones that with conventional geographical connection.

\begin{table}
\small
	\centering
	\caption{Results of the Three Spatial Durbin Models}\label{tab1}
    \begin{tabular}{p{1.7in}p{0.6in}p{0.6in}p{0.6in}} \hline
\textbf{Variable} & \textbf{Model 1} & \textbf{Model 2} & \textbf{Model 3} \\ \hline
\textbf{Local coefficients} &  &  &  \\
\textit{lnpcgdp} & -0.015 & 0.297${}^{*}$ & 0.124 \\
\textit{govexp} & 1.997${}^{*}$ & -3.046${}^{***}$ & -2.281${}^{*}$ \\
\textit{fdi} & 0.374 & -0.207 & -0.083 \\
\textit{edu} & -0.032${}^{***}$ & -0.031${}^{***}$ & -0.034${}^{***}$ \\
\textit{tradeopen} & 0.013 & 0.014${}^{*}$ & 0.016${}^{**}$ \\
\textit{lnfinload} & -0.113${}^{*}$ & -0.082 & -0.144${}^{***}$ \\
\textit{lnfixinvest} & 0.140${}^{**}$ & 0.159${}^{***}$ & 0.122${}^{*}$ \\
\textit{lnpoweruse} & 0.215 & 0.387${}^{*}$ & 0.314${}^{*}$ \\
\textbf{Local spatial coefficients} &  &  &  \\
\textit{W*lnpcgdp} & 0.291 & -0.408${}^{**}$ & -0.830${}^{***}$ \\
\textit{W*govexp} & -0.548 & 5.731${}^{**}$ & 1.910 \\
\textit{W*fdi} & 0.570 & 0.868${}^{**}$ & 1.310${}^{***}$ \\
\textit{W*edu} & -0.041${}^{*}$ & 0.016 & 0.008 \\
\textit{W*tradeopen} & -0.055 & -0.039${}^{*}$ & -0.056${}^{***}$ \\
\textit{W*lnfinload} & 0.341${}^{**}$ & -0.087 & -0.138${}^{***}$ \\
\textit{W*lnfixinvest} & 0.138 & -0.004 & 0.109 \\
\textit{W*lnpoweruse} & 0.932${}^{**}$ & -0.040 & 1.069${}^{***}$ \\
\textbf{Global spatial coefficient} &  &  &  \\ \hline
Spatial parameter ($\rho$) & -0.147 & 0.171${}^{**}$ & 0.327${}^{***}$ \\ \hline
\end{tabular}
\\Note: *, **, and *** denote statistically significant at 10\%, 5\% and 1\% level.
\end{table}

Notably, the coefficient values of the variables in Table \ref{tab1} cannot be directly explained. The marginal effect of each variable must be recalculated by considering the spatial effect. For Models 2 and 3, Table \ref{tab1} also highlights the presence of local spatial dependence. For Model 2, four variables lead to its local spatial dependence; among them, the coefficient values of $W*{\mathrm{ln} pcgdp\ }$ and $W*{\mathrm{ln} fixinvest\ }$ are negative, and those of $W*govexp$ and $W*fdi$ are positive. The results show that the high-tech industry agglomeration in a city is more likely to be boosted by neighboring cities with higher government expenditure and foreign direct investments.

However, the agglomeration effects might also be hampered by the dynamics of such neighboring cities with higher trade openness. One possible reason for this phenomenon might be that the cities with rapid economic development and a higher degree of openness have a relatively stronger appeal to the intellectual and capital resources required by high-tech industries, inhibiting the agglomeration of high-tech industries in the neighboring cities. Although the government's expenditure on science and technology and foreign direct investment may improve the city in many aspects such as the economy, people's livelihood and environment can also be a driver of the development of surrounding cities and increase the possibility of various resources flowing there and thus promote the high-tech industry agglomeration. Additionally, the attention of a local government to science and technology may stimulate the governments of neighboring cities to formulate similar incentive policies to promote the development of high-tech industries \cite{ref48}.

For Model 3, where cities have flow connection due to the migration of high-skilled workers, six variables result in local spatial dependence. The high-tech industry agglomeration is positively influenced by the improvement in FDI and power usage of the cities with high-skilled talent migration. The increase in the power usage of the whole society reflects the expansions of the population and industrial scale. On the one hand, these expansions need the support of high-tech industries, increasing the demand for the development of high-tech industries; on the other hand, the development of the city also increases the possibility that the resources required by the high-tech industry will be exported to surrounding cities. Therefore, it plays a leading role in the agglomeration of high-tech industries.

\subsection{Analysis of direct, indirect and total effects }
To discuss the impact of each influencing factor on industry agglomeration, we must calculate their marginal effect because the coefficients in Table \ref{tab1} do not consider the global spatial dependence of industry agglomeration. According to LeSage \textit{et al.} \cite{ref45}, the marginal effects of spatial econometric estimates can be divided into direct and indirect effects. In our study, for a city, the direct effect represents the impact on the industry agglomeration of a unit change in the independent variables in its own region. The indirect effect represents the impact on the industry agglomeration of a unit change in the independent variables in all other regions through neighbor relationships. Table \ref{tab2} reports the direct, indirect, and total effects of the independent variables of Models 2 and 3.

Model 2 considers the spatial effects caused by the proximity of geographic locations between cities. The total effect shows that the per capita GDP of the YRDUA region does not have a significant impact on the high-tech industry agglomeration, indicating that the high-tech industry agglomeration does not depend on the overall level of economic development of the YRDUA region. However, this insignificant impact is caused by significant negative indirect effects that counterbalance the positive direct effects. We must focus on the gap between a city and the neighboring cities, because the economic development of the city can promote high-tech industry agglomeration; however, it occurs at the expense of laggard neighboring cities. The reason for its occurrence might be that cities with faster economic development are more likely to attract the intellectual resources and capital necessary for high-tech industries, and thus, fewer resources flow into the neighboring cities, inhibiting industry agglomeration.

In terms of the share of government expenditure on science and technology, it has a strong negative effect on the high-tech industry agglomeration of the city, but a strong positive effect on that of neighboring cities, which is contrary to our expectation. The share of government expenditure on science and technology reflects the degree of government intervention on innovation. The higher the level of government intervention, the lower the efficiency of resource allocation. Therefore, excessive government intervention might be considered a low degree of marketization; additionally, it makes resources flow to cities with a higher degree of marketization and promotes the agglomeration of high-tech industries.

\begin{table}
\small
	\centering
	\caption{Direct, indirect, and total effects in Models 2 and 3}\label{tab2}
    \begin{tabular}{llll|lll} \hline
\textbf{} & \multicolumn{3}{p{1.7in}}{\textbf{Model 2}} & \multicolumn{3}{|p{1.7in}}{\textbf{Model 3}} \\ \hline
\textbf{} & \textbf{Direct} & \textbf{Indirect} & \textbf{Total} & \textbf{Direct} & \textbf{Indirect} & \textbf{Total} \\ \hline
\textit{lnpcgdp} & 0.279${}^{*}$ & -0.392${}^{*}$ & -0.112 & 0.102 & -1.169${}^{***}$ & -1.067${}^{***}$ \\
\textit{govexp} & -2.751${}^{***}$ & 5.910${}^{**}$ & 3.159 & -2.301${}^{*}$ & 1.666 & -0.635 \\
\textit{fdi} & -0.136 & 0.911${}^{***}$ & 0.775${}^{**}$ & -0.021 & 1.993${}^{***}$ & 1.973${}^{***}$ \\
\textit{edu} & -0.030${}^{***}$ & 0.013 & -0.017 & -0.034${}^{***}$ & -0.005 & -0.039${}^{*}$ \\
\textit{tradeopen} & 0.012 & -0.041${}^{**}$ & -0.029 & 0.014${}^{*}$ & -0.077${}^{**}$ & -0.063${}^{*}$ \\
\textit{lnfinload} & -0.083 & -0.109 & -0.192${}^{*}$ & -0.149${}^{***}$ & -0.284${}^{*}$ & -0.433${}^{**}$ \\
\textit{lnfixinvest} & 0.163${}^{***}$ & 0.027 & 0.189 & 0.126${}^{**}$ & 0.209 & 0.335 \\
\textit{lnpoweruse} & 0.366 & 0.025 & 0.391 & 0.342${}^{**}$ & 1.820${}^{***}$ & 2.161${}^{***}$ \\ \hline
\end{tabular}
\\Note: *, **, and *** denote statistically significant at 10\%, 5\% and 1\% level.
\end{table}

The proportion of inward foreign direct investments to GDP and the fixed asset investment in the whole society show significant indirect and direct effects, respectively. The development of high-tech industries requires more advanced technology and equipment; thus, the increase in the fixed asset investments in the whole society has promoted the agglomeration of the high-tech industries of the city. Additionally, the increase in the proportion of inward foreign direct investments over GDP increases the demand for high-tech industries and thus boosts their development in neighboring cities. The total loads from non-back financial intermediaries across YRDUA regions are observed to hamper the high-tech industry agglomeration across the entire region, indicating that most of the loans have flowed to other industries instead of high-tech industries. The local power usage is observed to have little role in industry agglomeration. However, after we consider population mobility, the results change substantially.

Model 3 considers the impact of spatial spillover effects caused by the migration of high-skilled talents between cities on high-tech industry agglomeration. Through our comparison of Models 2 and 3, we observe that the direct, indirect, and total effects of each factor in Model 3 are more significant. The results indicate that for high-tech industry agglomeration, the spatial spillover effects between the cities with the migration flow connection exists more obviously than that between the cities with a geographic connection. In terms of the total effect, FDI and electricity consumption play a pivotal role, and none of the per capita GDP, the provision of education quality, openness, or the local financial resources demonstrate positive effects.

In Model 3, the power usage of the whole society shows a very significant strong positive total effect. However, per capita GDP shows a very significant strong negative total effect. Both are decided by their significant indirect effects. The power usage of the whole society can reflect that the city has a larger industrial scale and population, that is, the city has a larger proportion of energy-intensive and labor-intensive industries, which is not advantageous to the development of high-tech industries. Therefore, the intellectual resources required for the knowledge-intensive and technology-intensive high-tech industries flow to other cities with the migration of high-skilled talents. The explanation of the indirect effect of per capita GDP is similar to that in Model 2. However, the marginal indirect effect of per capita GDP in Model 3 is much larger than that in Model 2, which means the spatial spillover effect is more significant between the cities with migration connection.

Compared with Model 2, the number of universities and openness degree also show negative total effects in Model 3; however, they have a different formation mechanism. The negative total effect of the number of universities is mainly because of the negative direct effect, indicating that for the cities in the YRDUA region, the large number of universities has not stimulated the agglomeration of high-tech industries. Although the university is a provider of intellectual resources, it is also a demander of intellectual resources from the perspective of research and development. Such a negative effect means that the high-tech enterprises and universities in the YRDUA region have not yet formed an effective cooperation but show a competitive relationship for local intellectual and capital resources. The negative total effect of the degree of trade openness is mainly caused by its negative indirect effect. Notably, this factor has some direct effect that is offset to some extent by its relatively stronger indirect effect. Thus, the high degree of trade openness of a city will be a driver of the development of high-tech industries of the city but inhibit the development of high-tech industries in other cities with population mobility.

\section{Conclusions}

The agglomeration of high-tech industries plays a critical role in increasing labor productivity and economic efficiency \cite{ref49}. It is recently found that the understanding of industrial agglomeration of high-tech industries can be enhanced from a spatial-spillover perspective. Admittedly, population mobility will bring about the innovation diffusion, which is a breeding environment for the development of high-tech industries, yet the spatial connection in the form of population flow has receive less attention.

In this paper, we focus on high-tech industries in the YRDUA region and explore the influencing factors of industrial agglomeration from a spatial-spillover perspective. The most important contribution of this paper to the literature is the validation of the spillover effect in a topological space formed by flows of high-skilled migrants between cities, which is formed by exploiting big data mining tools to 150,000 pertinent labor flows from Boss Zhipin, one of the largest online recruiting platforms in China. It breaks through the limitation that spillover effects are only discussed in neighboring regions. And it provides new view that even if the two regions are far away, their industrial agglomeration of high-tech industries may have spillover effects due to the migration of high-skilled talents, which is referred as the flow connection in this paper.

Specifically, this paper constructed three spatial econometric models based on three different mechanisms of connections, namely contiguity-based spatial weights, distance-boundary-based spatial weights, and weights based on the high-skilled migration flow, respectively, to examine and compare the spillovers of examined factors on the agglomeration of high-tech industries. Our regression results show that spatial spillover effects exist in the development of high-tech industries in the YRDUA region. Moreover, the spatial spillover effects are even stronger among cities with flow connections than among cities with stronger geographic connections.

In addition, we further investigated the sources of spillover effect. The proximate industries tend to cluster in a more open and market-oriented environment. Universities and enterprises currently are competing for resources in the research and development (R\&D) areas in high-tech manufacturing industries. Excessive government intervention is not conducive to the optimal allocation of resources, which might inhibit the development of high-tech industries. Even worse, one city's policies on science and technology are more likely to attract the attention of and be imitated by its neighboring cities. Under the premise of maintaining the stable development of the local economy, the government should improve the market environment to enable the development of more competitive high-tech industries. Additionally, enterprises, universities, and research institutes should be encouraged to cooperate, and the relevant policies that encourage the implementation of scientific research results should be further improved.

Notably, considering the existence of spatial spillover effects, improvements on a city's economic progress and openness degree hamper the high-tech industries in cities that are closely connected, and this inhibiting role is particularly stronger in cities with frequent immigration of high-skilled talents. In other words, more gains in foreign direct investments in a city causes the agglomeration of high-tech industries in other cities. This is partially due to the policy inertia when a city with higher uptake in traditional industries is less in favor of the energy-efficient and labor-saving high-tech industries. Consequently, the intellectual resources required for knowledge-intensive and technology-intensive industries tend to flow to the neighboring cities or the cities with the migration of high-skilled talents.



\end{document}